\documentstyle[prb,twocolumn,epsfig,aps]{revtex}
\begin{document}

\draft
\preprint{}


\title{ A quantum generalization of the thermal viscous friction law}
\author{P. Shiktorov, E. Starikov, V. Gru\v zinskis}
\address{Semiconductor Physics Institute,
A. Go\v stauto 11, 2600 Vilnius, Lithuania}
\author{L. Reggiani}
\address{INFM - National Nanotechnology Laboratory,
Dipartimento di Ingegneria dell' Innovazione,
Universit\`a di Lecce, Via Arnesano s/n, 73100 Lecce, Italy}
\date{\today}
\maketitle
\begin{abstract}
On the basis of the equivalence of the energy balance description
at micro- and macro-level we propose a quantum generalization of
the viscous friction law for a macroscopic Langevin equation
describing thermal fluctuations without the zero point
contribution. This equation recovers the classical case in the
limit of $h \rightarrow 0$. 
In particular it satisfies the quantum regression theorem and  
resolves several anomalies appearing in the quantum extension 
of the fluctuation dissipation theorem.
\end{abstract}
\pacs{PACS numbers: 03.65.Ca, 05.30.-d, 05.40.+j}
\narrowtext
The classical Langevin equation
for the damped harmonic oscillator with eigenfrequency $\omega_s$
in the presence of friction and of a stochastic Langevin force
$f_C(t)$ in the time domain writes:
\begin{equation}
{d^2 \over dt^2} x  + \gamma {d \over dt} x + \omega_s^2 x  = f_C(t)
\label{1}
\end{equation}
Here $\gamma$ is the viscous friction coefficient and the
correlation function of the thermal Langevin force is given by
$<f_C(0)f_C(t)> = 2kT \gamma \delta(t)$. Equation (1) forms a
basis for the fluctuation dissipation theorem (FDT) in its
classical form:
\begin{equation}
S^C_{xx}(\omega) = {2kT \over \omega} Im\{\alpha^C_x(\omega)\}
\label{2}
\end{equation}
with $S^C_{xx}$ the two-side spectral density of the displacement
fluctuations in the frequency range $[-\infty , \infty ]$ and
$\alpha^C_x(\omega)$ the classical susceptibility (i.e. the
Fourier transform of the response function) that for the
damped harmonic oscillator writes:
\begin{equation}
\alpha^C_x(\omega) = {1 \over \omega_s^2 - \omega^2 - i \gamma
\omega } \label{3}
\end{equation}
\par
According to a recent comment of van Kampen\cite{kampen01}: {\it
the shortcoming of the above Langevin scheme is that it cannot be
extended to quantum mechanics; attempts have been made to simple
generalize  Eq. (\ref{1})  so as to turn it into an equation of
operators, but that did not work.\cite{benguria80}} Indeed, one
way to perform such a generalization is to introduced a relaxation
already at the level of  the Heisenberg equations of motion for
operators, the so called {\it quantum Langevin equations}.
\cite{{gardiner81,meystre91}} 
However, as mentioned by van Kampen, 
such an approach is internally contradictory since
formally it implies that a set of quantum-mechanical operators and
their commutation rules which describe the physical system are relaxing.
This would be a severe 
violation of the laws of quantum mechanics
since commutation relations must be valid at all times.\cite{meystre91}

To overcome this contradiction, a second way of proceeding is to
introduce a relaxation for observables only.
Accordingly, relaxation
appears just in the macroscopic equations of motion for the
observables, such as $x=Tr\{\hat\rho\hat x\}$, that are obtained after
averaging over the statistical operator
$\hat\rho$.\cite{meystre91,grabert84} 
This second approach is free from
the above contradiction since the quantum corrections to the
Langevin scheme are formulated  only at the macrolevel on the
basis of the quantum  FDT (QFDT) given usually
by the well known  Callen
and Welton\cite{callen51} expression. In full analogy with the
classical case, here the friction remains viscous and the
modifications concern with the Langevin force spectrum only.
However, some contradictions appear in this scheme too. For
example, when  thermal fluctuations are considered this second
approach enters in conflict with the quantum regression theorem
(QRT). 
Usually such a conflict is announced in the form 
{\it "there is no quantum
regression theorem"} \cite{talkner86,ford96}, despite of the fact
that the QRT was derived by Lax\cite{lax63} independently from the
QFDT.
\par
The aim of this letter is to remove these contradictions
of the second approach by
introducing a quantum viscous force law
which generalizes the relaxation of observables given by Eq. (1)
to the quantum case. 
To this purpose: (i) the classical viscous
friction law is replaced by the quantum analog as
\begin{equation}
\gamma {d \over dt} x
\ \ \rightarrow \ \
\gamma {2 \over \beta} sin({\beta \over 2} 
{d \over dt}) x
\label{4}
\end{equation}
and (ii) the classical Langevin force correlator by the quantum
analog:
\begin{equation}
<f_Q(0)f_Q(t)> = {2\hbar \gamma \over \pi \beta}
\int_0^{\infty} exp(- {\beta \omega \over 2})cos(\omega t)
d \omega
\label{5}
\end{equation}
where $1/\beta = kT/\hbar$.
\par
The above quantum generalization of the Langevin scheme satisfies
the QFDT  in the form originally proposed by
Nyquist\cite{nyquist28} without the contribution of the zero-point
energy of the thermal bath radiation: 
\begin{equation}
S^Q_{xx}(\omega)=2\hbar sgn\{\omega\} \overline{N}_T(\omega) Im\{\alpha^Q_x(\omega)\}
\label{6}
\end{equation}
with the quantum susceptibility $\alpha^Q_x(\omega)$:
\begin{equation}
\alpha^Q_x(\omega) = {1 \over {\omega_s^2 - \omega^2 - i{ \gamma \over \beta} sinh ({\omega \beta \over 2})}}
\label{7}
\end{equation}
Here
%
$
\overline{N}_T(\omega)=[exp(\beta|\omega|)-1]^{-1}
$
%
is the average number of thermal photons with frequency $\omega$.
\par
Below, we shall demonstrate that the above extended approach
is consistent with the principle of detailed energy balance, while
the conventional approach based on
the classical viscous force and Callen-Welton
expression\cite{callen51} of the QFDT violate this principle.

{\it Micro- and macro-level of detailed energy balance
description.} We shall consider a sufficiently large isolated
system, with Hamiltonian $\hat H=\hat H_S+\hat H_T+\hat V$, which
can be decomposed into two interacting subsystems. The first one
is described by $\hat H_S$ and corresponds to the physical system
under test. The second subsystem is described by $\hat H_T$ and
represents the surrounding world. The interaction between these
subsystems is described by the Hamiltonian $\hat V=-\hat x \hat
f$, where $\hat x$ and $\hat f$ are operators representing
dynamical variables pertaining to the S- and T-subsystems,
respectively.

By using the standard procedure\cite{meystre91} to
go from the density matrix of the whole system $\hat \rho$ to the
reduced density matrixes of subsystems $\hat \rho_S = Tr_T\{\hat
\rho\}$ and $ \hat \rho_T = Tr_S\{\hat \rho\}$
in the framework of the energy representation of $\hat \rho_i =
\hat \rho_i (\hat H_i)$ one obtains the following equation for the
time evolution of the average energy $<\hat H_i>$ in the $i$-th
subsystem ($i=S,T$):
%
%
$${d \over dt} \biggl [{<\hat H_T> \atop <\hat H_S>}
\biggr ]= {\pi \over \hbar} \sum_{m,n}^S \ \sum_{M,N}^T \
|x_{mn}|^2 |f_{MN}|^2
$$
\begin{equation}
\times
\biggl [{ \omega_{mn}^S \atop \omega_{MN}^T}\biggr ]
( \rho_m^S\rho_M^T - \rho_n^S\rho_N^T)
\delta(\omega_{mn}^S + \omega_{MN}^T)
\label{8}
\end{equation}
where $\rho_m^i$ is the probability to find the $i$-th subsystem
in the state with eigenenergy $E_m^i$,
$\omega_{mn}^i=(E_m^i-E_n^i)/\hbar$ the frequency of permitted
transitions, $x_{mn}$ and $f_{MN}$ the matrix elements of the
$\hat x$ and $\hat f$ operators, respectively. 
When $<\hat H_i>=const$, 
from Eq. (8) one directly obtains the condition of
microscopic detailed energy balance\cite{kampen92} 
\begin{equation}
\rho_m^S \rho_M^T = \rho_n^S \rho_N^T   \ \ \ \ \
\end{equation}
\par
To formulate the conditions of energy balance at the macroscopic
level of description, we replace the factor $\delta(\omega_{mn}^S +
\omega_{MN}^T)$ by $\int \delta(\omega_{mn}^S -\omega) \delta
(\omega_{MN}^T + \omega) d\omega $ and by using  the definition of
spectral density given by:
\begin{equation}
J_{ii}(\pm \omega)=
2\pi
\sum_{m,n} \rho_n |i_{mn}|^2
\delta (\omega_{mn} \mp \omega)
\end{equation}
which corresponds to the quantum correlation function 
$Tr\{\hat\rho \hat i(\pm \tau)\hat i(0)\}$ of the dynamical 
operators $\hat i=\hat x,\hat f$, we rewrite Eq. (8) as:
%
$$
{d \over dt} <\hat H_T> =
- {d \over dt} <\hat H_S>=
$$
\begin{equation}
  {1 \over 4\pi  \hbar}
\int_{-\infty}^{\infty} \omega [J_{xx}(-\omega)J_{ff}(\omega)
- J_{xx}(\omega) J_{ff}(-\omega)] d \omega
\end{equation}
From Eq. (11) we obtain
the condition of macroscopic detailed energy balance
(MaDEB) as:
%
\begin{equation}
J_{xx}(-\omega)J_{ff}(\omega) = J_{xx}(\omega) J_{ff}(-\omega)
\end{equation}
which requires to be fulfilled for any value of the current
frequency $\omega$. We notice that the condition given by Eq. (12)
is not the only form in which the MaDEB can be expressed. By using
the expression for the imaginary part of the macroscopic
susceptibilty\cite{zubarev71,kubo85}:
\begin{equation}
Im\{\alpha_{i}(\omega)\} =
{1\over 2\hbar} [J_{ii}(\omega) - J_{ii}(-\omega)]
\label{13}
\end{equation}
which describes the dissipative part of the response 
of one subsystem to the action of the
other one (for example, if $i=x$ is the response then the action
is given by $f=Tr\{\hat \rho_T \hat f\}$ and {\it vice versa}) the
MaDEB conditions can be rewritten in an equivalent form as:
%
\begin{equation}
Y_{x}^{\mu}(\omega)Im\{\alpha_f(\omega)\} =
Y_{f}^{\mu}(\omega)Im\{\alpha_x(\omega)\}
\end{equation}
where, in general,  the weighted symmetric spectral 
density $Y_{i}^{\mu}(\omega)$ is given by:
%
\begin{equation}
Y_{i}^{\mu}(\omega)=(1-\mu)J_{ii}(-|\omega|)+\mu J_{ii}(|\omega|)
\end{equation}
and the weighting factor $\mu$ 
can be treated as an arbitrary parameter.
Note, that to
derive Eq. (14) one need to preserve the antisymmetric property of
the term in the square brackets of Eq. (11) under all the
transformations. Since by definition $Im\{\alpha_i(\omega)\}$ is
an antisymmetric function of frequency, to keep the antisymmetry
of the whole expression, $Y_{i}^{\mu}(\omega)$  must be treated as
a symmetric function, i.e. as a function of the absolute value of
frequency.
\par
In the following we shall consider the formulation of MaDEB
conditions by using the ratios of some characteristics of only one
of the two subsystems. In such a representation, Eq. (12) takes
the form:
%
\begin{equation}
{J_{xx}(-\omega)\over J_{xx}(\omega) } =
{J_{ff}(-\omega)\over J_{ff}(\omega) }
\equiv p(\omega)
 \ \ or \ \
{J_{ii}(-\omega)= p(\omega)  J_{ii}(\omega) }
\end{equation}
where $p(\omega)$ is a factor common to both subsystems
satisfying the relation $p(-\omega)=p^{-1}(\omega)$. The MaDEB
condition given by Eq. (14) can be represented in
the analogous forms:
%
\begin{equation}
{Y_{i}^{\mu}(\omega) = g^{\mu}(\omega) Im\{\alpha_i(\omega)\} }
\end{equation}
where
%
\begin{equation}
g^{\mu}(\omega)=2\hbar sgn\{\omega\}
\biggl[
{p(|\omega|)\over 1-p(|\omega)|} + \mu
\biggr]
\end{equation}
Here, $g^{\mu}(\omega)$ depends on 
$\mu$
and is determined by the frequency dependence of $p(\omega)$ 
at $\omega > 0$ only.

When the physical system interacts with the radiation bath 
described by a certain, not obligatory thermal,
distribution of photon numbers in the radiation modes,  
the bath spectral densities  $J_{ff}(\pm \omega)$
can be represented as\cite{loudon73}:
%
\begin{equation}
{J_{ff}(\pm |\omega)|}
=
2\pi{G(\omega)\gamma^2(\omega)\hbar \over \omega }
\biggl [
{ \overline{N}(\omega) +{1\over 2} \pm {1\over 2}}
\biggr ]
\end{equation}
where 
$G(\omega)$ is the density of radiation modes in the bath and
$\gamma^2(\omega)$ the electro-dipole matrix element square. 
Substitution of Eq. (19) into the MaDEB condition given by Eq. (16) 
allows us to rewrite the latter in terms of
the average number of photons 
in the radiation mode with frequency $\omega$:
%
\begin{equation}
\overline{N}(\omega)\equiv Tr\{\hat\rho_T\hat n_{\omega}\}
={p(|\omega|)\over 1-p(|\omega|)}
\end{equation}
Accordingly, the MaDEB condition given by Eq. (17) takes the form:
%
\begin{equation}
{Y_{x}^{\mu}(\omega)}
=2\hbar sgn\{\omega\} [\overline{N}(\omega)+\mu]
 {Im\{\alpha_x(\omega)\}}
\end{equation}

By using Eq. (21), let us give an interpretation to the
energy balance at the macro-level.
As follows from Eq. (14) the energy balance description at the 
macro-level 
is formulated only in the framework of the notion of a classical absorption 
(i.e. the energy dissipation by one subsystem from another) 
which is described by $Im\{\alpha_i(\omega)\}$. 
With respect to each
interacting subsystem such an energy transfer occurs in one direction,
from outside to inside. 
The opposite process, which returns the
energy back, i.e. the energy emitted by the subsystem is absent.
This representation is due to the definition of $\alpha_i(\omega)$
as the response of the system to a classical external
force which, by definition, does not take into account the process
of spontaneous emission. 
Therefore,  with respect to a single
subsystem, one side of Eq. (14) must be treated as the absorbed
(dissipated) power and the other side as the emitted (returned)
power. 
Indeed, a direct substitution of Eq. (19) into Eq. (14) 
gives us again Eq. (21).  Here, the right-hand side, which contains
$Im\{\alpha_x(\omega)\}$, describes the power gained from the
radiation bath and dissipated in the system. 
The left-hand side, which is given identically by $Y_x^{\mu}(\omega)$, 
describes the power returned by the system to the bath. 
However, as follows from Eq. (21), at the
MaDEB level the state of the bath radiation 
which is proportional to
$(\overline{N}(\omega)+\mu)$ and the spectrum of fluctuations
$Y_x^{\mu}(\omega)$ are formally uncertain since $\mu$ can take
arbitrary values. 

{\it Equivalence between  MiDEB and MaDEB.}
As follows from the above consideration,
micro- and macro-levels of the energy balance description
can not be treated as entirely equivalent. 
Indeed, the former is formulated in terms of statistical operators
$\hat \rho_i$ (see Eq. (9)),
which describe the interacting subsystem microstates only.
By  contrast, the latter is based on characteristics 
which are by definition invariant with respect to these microstates
since all the MaDEB conditions 
are formulated for the quantities 
statistically averaged over $\hat \rho_i$.
Moreover, the MaDEB conditions can include the extra arbitrary parameter
$\mu$ which does not directly follow from MiDEB. 

It is easy to see, that
the energy balance descriptions based on Eqs. (8) and (11)
will be equivalent if the following relation is satisfied
at all the frequencies of permitted transitions between the subsystems:
%
\begin{equation}
{\rho_m^S \over \rho_n^S} = {\rho_N^T \over \rho_M^T} =
p(\omega)\biggl |_{\omega=\omega_{mn}=\omega_{NM}}
\end{equation}
At thermal equilibrium, when $\hat \rho_i\sim exp(-\hat H_i/kT)$,
the common factor $p(\omega)=exp(-\hbar\omega /kT)$ becomes
the universal function of the temperature and frequency only,
so that the MaDEB condition given by Eq. (21)
at $\mu=0$ and $1/2$ coincides with the Nyquist\cite{nyquist28} 
and Callen-Welton\cite{callen51} versions
of the QFDT, respectively.
{\it This fact allow us
to give to the QFDT an alternative physical interpretation with
respect to the conventional one as a relation which describes the
MaDEB between two interacting physical systems under thermal
equilibrium.}

The explicit presence of the uncertainty
introduced by $\mu\neq 0$ in the MaDEB given by Eqs. (17) and (21)
leads to a disagreement between
the micro- and the macro-level of description since, in essence,
it claims that the statistical and photon-number operators, 
$\hat\rho_T$ and $\hat n$, respectively,  are not sufficient to determine
the average number of photons in the mode, i.e. 
$\overline{N}\neq Tr\{\hat \rho_T \hat n\}$.
In accordance with the MiDEB and MaDEB equivalence condition given by
Eq. (22), $\mu$ can  always be neglected by
renormalizing  the average number of photons in the bath
$N'(\omega)=\overline{N}(\omega)+\mu$.
Such a renormalization
will change only the value of $\hat \rho_S$, i.e. a
microstate, by transforming $Y_x^{\mu}(\omega)$ to
$J_{xx}(-|\omega)|$. 
In this case, only the photon part,
$\hbar\omega\overline{N}(\omega)$, of the full energy of the field
is involved in the dissipation process described by
$Im\{\alpha_x(\omega)\}$, and the process of returning back the
energy is accomplished by the spontaneous emission characterized
by the spectral density $J_{xx}(-|\omega|)$. 
Thus, when $\mu =0$, Eq. (21) describes
the energy exchange between the two subsystems consistently with
the microscopic picture.

As follows from Eq. (22), in the frequency regions where
$p(\omega)=1$ the energy transitions are under saturation, that is
$\rho_m=\rho_n$ at $\omega=\omega_{mn}$. 
The saturated transitions
can only appear jointly in both the interacting subsystems.
In the opposite case, the MiDEB conditions are violated. 
By using Eq. (13) and Eq. (16), 
the imaginary part of the susceptibilities can be represented as:
%
\begin{equation}
Im\{\alpha_{i}(\omega)\} =
{1\over 2\hbar} [1-p(\omega)] J_{ii}(\omega)
\end{equation}
As follows from Eq. (23), 
at the frequencies of the saturated transitions the macroscopic response 
of both the subsystems must
not contain the dissipative component, since
$Im\{\alpha_i(\omega)\}=0$ at $p(\omega)=1$. 

If one of the subsystems is the radiation bath, 
then, as follows from Eq. (20), the dissipativeless interaction 
occurs at all the frequencies where  
$\overline{N}(\omega)\rightarrow \infty$. 
Note, that the state of the radiation
in these modes is similar to the classical state with exactly
determined phase\cite{loudon73}.
Under thermal equilibium the frequencies of
the saturated transitions are determined by the condition
$exp(-\beta\omega)=1$ and they correspond to the so-called
Matsubara frequencies $\omega=\Omega_k\equiv i2\pi\beta^{-1}k$
with $k=0,\ \pm 1,\ \pm 2,\ ...$ placed at $k\neq 0$ 
in the imaginary axis.
The classical viscous friction law 
introduced usually to describe the macroscopic dissipation
satisfies
the MiDEB conditions of the dissipativeless interaction,
however, on the real frequency axis only.
Indeed on the real axis
$\overline{N}(\omega)\rightarrow \infty$
only at $\omega=0$ where the viscous friction law
always leads to $Im\{\alpha_x(\omega)\}=0$. 
However, this law fails by extendinding these MiDEB requirements 
to the whole complex frequency plane since it
formally violates the energy balance at the Matsubara frequencies
with $k\neq 0$ where $Im\{\alpha_x(\Omega_k)\}\neq 0$. 
Such a violation is just the source of the QRT-QFDT
conflict\cite{talkner86,ford96} and of unremovable divergences of
the correlation functions of fluctuations when $T\rightarrow
0$.\cite{gardiner81,grabert84}

{\it The Langevin scheme}. 
In the framework of linear response theory, 
the spectral densities of fluctuations of the observable 
and the external force by which they are initiated
are interrelated by the simple relation:
%
\begin{equation}
Y_{x}^{\mu}(\omega)=
|\alpha_x(\omega)|^2Y_{f}^{\mu}(\omega)
\end{equation}
This allows us to rewrite MaDEB condition given by Eq. (17) in the form:
%
\begin{equation}
{Y_{f}^{\mu}(\omega) = - g^{\mu}(\omega) Im\{\alpha_x^{-1}(\omega)\} }
\end{equation}
where 
$Im\{\alpha_x^{-1}(\omega)\}=-Im\{\alpha_i(\omega)\}
/|\alpha_i(\omega)|^2$ 
describes the relaxation law of fluctuations of the observable.
For instance,  $Im\{\alpha_x^{-1}(\omega)\}=-\gamma\omega$
in the case of the classical viscous friction. 
In the framework of the Langevin approach
the quantity $Y_{f}^{\mu}(\omega)$ determines
the spectral density of the Langevin force which initiates
fluctuations of the observable described by the spectral density
$Y_{x}^{\mu}(\omega)$.

To match the Langevin scheme with MiDEB it is necessary:

(i) To exclude 
all the extra contributions caused by the free parameter $\mu$.
It means that the spectrum of fluctuations of the observable
 must coincide with 
the spontaneous emission spectrum of the system under test.
This MiDEB requirement agrees with the opinion often meet 
in the literature\cite{henry96,gavish00} that 
zero-point fluctuations
which corresponds to MaDEB condition with $\mu=1/2$ 
cannot be directly detected.

(ii) To satisfy the dissipativeless interaction requirement,
the condition $Im\{\alpha_x(\omega)\}=0$ must hold
in all the points of the complex frequency plane where $p(\omega)=1$.

The former requirement is satisfied in a trivial way by assuming $\mu=0$.
One of the possible ways to satisfy the latter one
is to modify the viscous friction law.
For this sake, by taking into account that by definition 
$Im\{\alpha_x^{-1}(\omega)\}$ and $Y_f^0(\omega)$
are  odd and even functions of frequency, respectively,
let us rewrite Eq. (25), which determines the specral density
of the Langevin force, in the form:
%
$$
Y_f^0(\omega)=
\biggl[
2\hbar{\gamma \over \beta}
e^{-{\beta|\omega| \over 2}} 
\biggr]
\times
\biggl[
{\beta \over \gamma} 
{ Im\{\alpha_x^{-1}(\omega)\} \over
{(e^{-{\beta\omega \over 2}} - e^{{\beta\omega \over 2}}) } 
}
\biggr]
\eqno(26)
$$
In the classical limit $\hbar \rightarrow 0$,
the first factor in the r.h.s. of Eq. (26)
gives the classical spectral density of the Langevin force,
$2kT\gamma$, while the second factor is identically equal to unity
for the case of a classical viscous friction,
i.e. when $Im\{\alpha_x^{-1}(\omega)\}=-\gamma\omega$.
By assuming that the second factor keeps the same value
also at $\hbar \neq 0$,
one obtains the quantum generalization
of the thermal law of viscous friction in spectral representation as:
%
$$
 Im\{\alpha_x^{-1}(\omega)\} = {\gamma \over \beta}
{(e^{-{\beta\omega \over 2}} - e^{{\beta\omega \over 2}}) } 
\eqno(27)
$$
which satisfies the second requirement of MiDEB
in the whole complex frequency plane.
The corresponding time representation is given by expression (4).
In so doing, the first factor in the r.h.s of Eq. (25)
gives the quantum analog of the spectral density of the 
thermal Langevin force 
which corresponds to the correlation function given by Eq. (5).
 
In conclusion,  we have proposed a thermal
quantum viscous friction law at a macroscopic level.
This law leads to a macroscopic quantum Langevin equation 
that does not include the zero point contribution in 
the spectral density of the fluctuating observable
under thermal equilibrium conditions.
The most relevant implications of this generalization are:
(i) in the classical limit $h \rightarrow 0$ the quantum Langevin 
equation recovers the classical one as it should;
(ii) it resolves several anomalies appearing in the quantum
extentions of the fluctuation dissipation and regression theorems,
such as the well known QRT-QFDT conflict, and various divergencies
originated by the Callen-Welton form of the QFDT;
(iii) the prediction of an exponential like
decay of the susceptibility at $\omega \gg kT/\hbar$.

{\it Acknowledgments.}
Partial support from European Commission through project 
No. IST2001-38899 is gratefully acknowledged.

\end{document}